\documentclass[namedreferences]{SolarPhysics}
\usepackage[optionalrh]{spr-sola-addons} 
\usepackage{graphicx}        
\usepackage{color}           
\usepackage{url}             
\usepackage{rotating}
\usepackage{enumerate}

\newcommand{\aap}{    {\it Astron. Astrophys.}}

\newcommand{\apj}{    {\it Astrophys. J.}}

\newcommand{\solphys}{{\it Solar Phys.}}

\newcommand{\ssr}{    {\it Space Sci. Rev.}}

\begin{document}

\begin{article}

\begin{opening}

\title{Quasi-periodic Variations in the Hard X-ray emission
 of a Large Arcade Flare}

%
\author{J.~\surname{Jakimiec}\sep
        M.~\surname{Tomczak}
       }

%
\runningauthor{J.\,Jakimiec \& M.\,Tomczak} \runningtitle{QPO in
HXRs of an arcade flare}

%
  \institute{Astronomical Institute, University of Wroc{\l }aw,
  ul. Kopernika 11, 51-622 Wroc{\l }aw, Poland,
                     email: \url{jjakim; tomczak@astro.uni.wroc.pl}\\
             }

\begin{abstract}
Quasi-periodic oscillations of the hard X-ray (HXR) emission of the large flare of 2 November 1991 have been investigated using HXR light curves and soft X-ray and HXR images recorded by the {\sl Yohkoh} X-ray telescopes. The results of the analysis of these observations are the following: i) The observations confirm that electrons are accelerated in oscillating magnetic traps which are contained within the cusp magnetic structure. ii) The amplitude of the HXR pulses increase due to the increase in the amplitude of the magnetic trap oscillations and the increase in the density within the traps caused by the chromospheric evaporation upflow. iii) The increase in the amplitude of the HXR pulses terminates when further increase in the density inside the traps inhibits the acceleration of electrons. iv) The model of oscillating magnetic traps is able to explain time variation of the electron precipitation, strong asymmetry in precipitation of accelerated electrons, and systematic differences in the precipitation of 15 and 25 keV electrons. v) We have obtained a direct observational evidence that strong HXR pulses are the result of the inflow of dense plasma coming from the chromospheric evaporation, into the acceleration volume.
\end{abstract}
%

\end{opening}

\section{Introduction}
     \label{intr}

In the hard X-ray (HXR) emission of many flares quasi-periodic variations were observed (\opencite{lip78}; see also the review of \inlinecite{n+m09} and references therein).

In our previous papers (\opencite{paper1} (Paper I), \citeyear{paper2} (Paper II)) we investigated flares with periods $P = 10-60$~s, but in Paper I we have found three flares with periods $P > 120$~s. They turned out to be large arcade flares. Investigation of the quasi-periodic oscillations in such large flares is very important, since their large sizes allow us to investigate the structure of the oscillation volume more comprehensively. Unfortunately, appropriate observations of the X-ray oscillations in such large flares are very rare. In the present paper we investigate an interesting example of such a large flare of 2 November 1991. Section~\ref{obs} contains the analysis of observations and Sections \ref{disc} and \ref{sum} present discussion and summary.

\begin{figure}
\centerline{\includegraphics[width=0.8\textwidth]{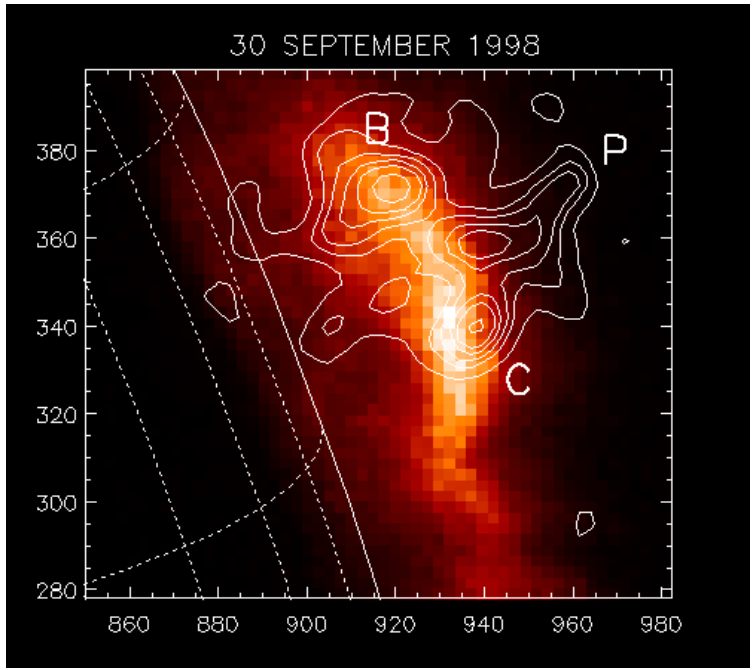}}
\hfill \caption{Soft X-ray image of the 30 September 1998 flare
(intensity scale in colour). The isocontours show the HXR 14-23 keV intensity. The solid line shows the solar limb and dashed lines are on the solar disc.} \label{sep30}
\end{figure}

In Figure~\ref{sep30} we reproduce the X-ray image of the large flare of 30 September 1998 (see Paper I), since it will be helpful in our analysis of the 2 November 1991 flare. In the colour SXR image a long arcade channel is seen. In the HXR image (isocontours) a triangular (``cusp'') structure is seen above the arcade channel. The strong HXR sources, B and C, occur at the places where the cusp structure contacts with the arcade channel. This suggests that there is magnetic connection between the cusp and the channel at B and C, which allows the electrons accelerated in the cusp to penetrate into the dense plasma of the channel and to generate the enhanced HXR emission there.

\section{Observations and Their Analysis}\label{obs}

\begin{figure}
\centerline{\includegraphics[width=1\textwidth]{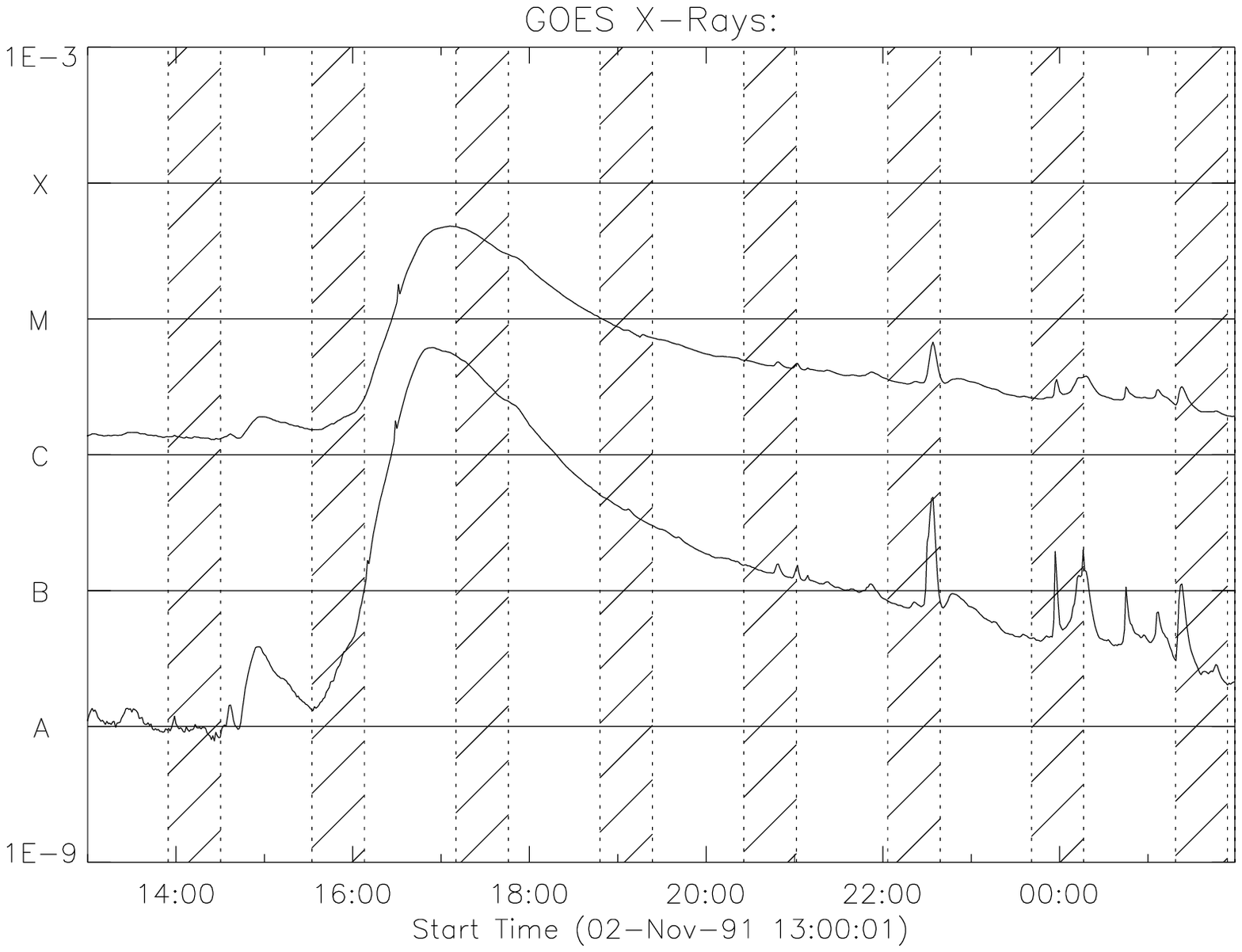}}
\hfill \caption{The standard GOES light curves for the flare of 2 November 1991 (upper curve -- 1-8 \AA\ range, lower curve -- 0.5-4 \AA\ range). The hatched areas show the {\sl Yohkoh} satellite nights.} \label{goes}
\end{figure}

In the present paper we investigate a large arcade flare which occurred at the western limb on 2 November 1991 (see X-ray images in Figures \ref{hxt1}--\ref{sxt} and \ref{hxt3}--\ref{hxt4}). It was a long duration event (LDE) of GOES class M4.8. The soft X-ray emission had begun to rise at 15:32 UT and it reached its maximum at about 17:00 UT (see the GOES light curves in Figure~\ref{goes}), {\sl i.e.} it was of ``slow-LDE'' type according to \inlinecite{h+m00}. The H$\alpha$ emission began at the northern footpoint of the flaring loop at 16:19 UT and it was seen up to 18:10 UT (see {\sl Solar-Geophysical Data} No.~573/II).

\begin{figure}
\centerline{\includegraphics[width=0.75\textwidth]{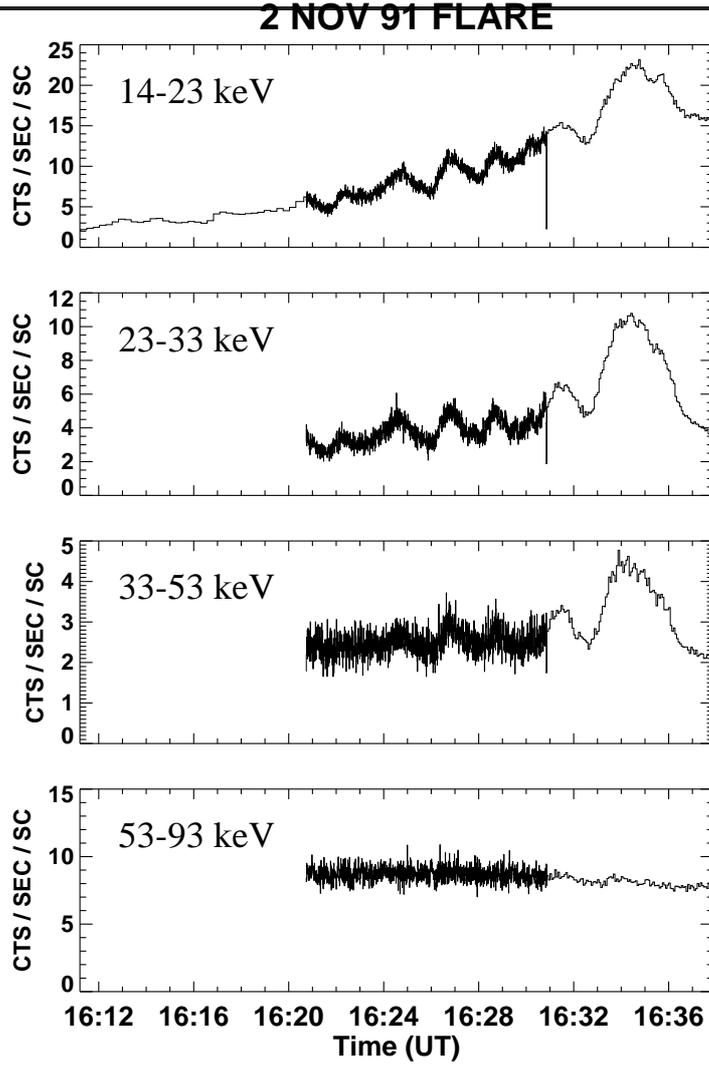}}
\hfill \caption{The {\sl Yohkoh} hard X-ray (HXR) light curves in four energy ranges.} \label{lc-hxt}
\end{figure}

\begin{figure}
\centerline{\includegraphics[width=1\textwidth]{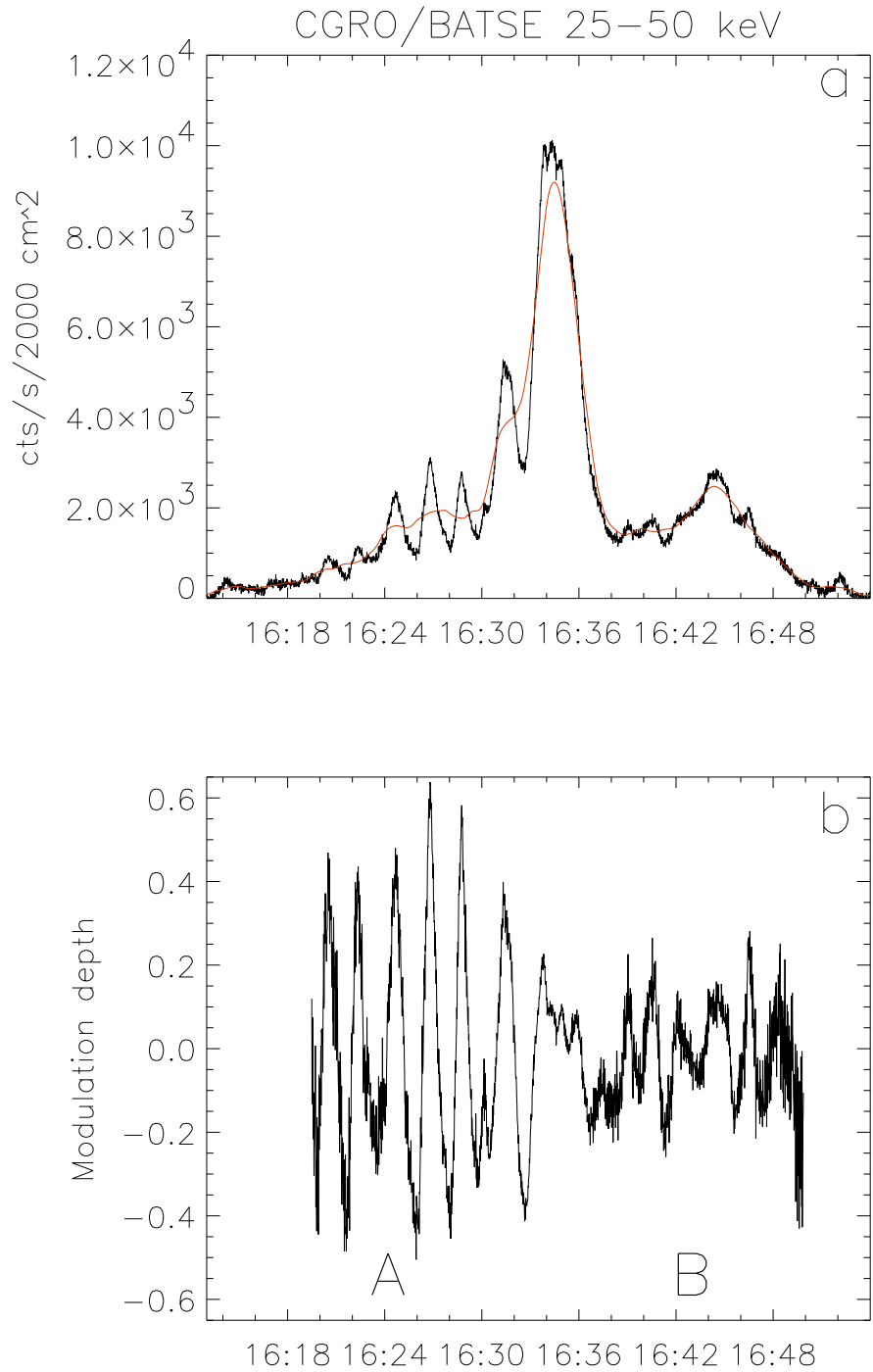}}
\hfill \caption{a) The {\sl Compton Gamma Ray Observatory} HXR light curves. The red, smoother, curve is the running mean of the original light curve. b) The normalized light curve, $S(t)$ [see Equation (1)].} \label{batse}
\end{figure}

The hard X-ray light curves, recorded by the {\sl Yohkoh} Hard X-ray Telescope, HXT \cite{kos91} and the {\sl Compton Gamma Ray Observatory Burst and Transient Source Experiment}, BATSE \cite{fis92}, are shown in Figures \ref{lc-hxt} and \ref{batse}a. The nominal energy range of the BATSE observations is $h{\nu} > 25$~keV. The comparison of the light curves in Figures \ref{lc-hxt} and \ref{batse} shows that the strong peak at about 16:34 UT is more dominant in the BATSE observations than in the {\sl Yohkoh} 23-33~keV energy range. We have compared the ratio of the amplitude of the strong peak and of the weak peaks (16:21--16:30 UT) in these sequences and we have found that the BATSE ratio is in good agreement with the {\sl Yohkoh} 33-53~keV ratio, which indicates that the actual energy range of the BATSE observations was $h{\nu} > 33$~keV.

\subsection{Analysis of Quasi-periodicity of the HXR Pulses}\label{qper}

To separate the HXR pulses from the smooth HXR rise we have calculated the normalized time series $S(t)$(see Paper II).
\begin{equation}
S(t) = \frac{F(t) - \hat{F}(t)}{\hat{F}(t)},
\end{equation}
where $F(t)$ is the measured HXR flux and $\hat{F}(t)$ is the running average of $F(t)$. The red curve in Figure \ref{batse}a shows $\hat{F}(t)$ calculated with averaging time $\delta{t} = 120$\,s. The normalized time series, $S(t)$, is shown in Figure \ref{batse}b. Our basic method to recognize quasi-periodic sequences is the following. We measure time-intervals, $P_i$, between successive HXR peaks and calculate the period, $P = <P_i>$, and its standard (r.m.s.) deviation, ${\sigma}(P)$. Our criterion of a quasi-periodicity is ${\sigma}(P)/P \ll 1$. The values of $P$ and ${\sigma}(P)$ for different parts of the impulsive phase of the investigated flare are given in Table~\ref{tab1}.

\begin{table}
 \caption{Parameters of the quasi-periodic oscillations of the 2 November 1991 flare}
 \label{tab1}
\begin{tabular}{cccccc}
\hline \multicolumn{2}{c}{During impulsive} & \multicolumn{2}{c}{At the maximum} & \multicolumn{2}{c}{After impulsive} \\
\multicolumn{2}{c}{phase rise} & \multicolumn{2}{c}{of impulsive phase} & \multicolumn{2}{c}{phase} \\
\multicolumn{2}{c}{16:20--16:30 UT} & \multicolumn{2}{c}{16:30--16:36 UT} & \multicolumn{2}{c}{16:38--16:49 UT} \\
 \hline
 $P$ [s] & Amp $S$ & $P$ [s] & Amp $S$ & $P$ [s] & Amp $S$ \\
 & & & & & \\
129$\pm$3 & 0.94 & 164 & 0.67 & 115$\pm$10 & 0.38 \\
 \hline
\end{tabular}
\begin{list}{}{}
\item $P$ is the mean time-interval between successive HXR peaks, and \\ Amp $S$ is the mean value of full amplitude of the $S$-function \\ measured for individual HXR peaks.
\end{list}
\end{table}

Figures \ref{lc-hxt} and \ref{batse}, together with Table~\ref{tab1}, show clear quasi-periodicity of the HXR pulses during the impulsive phase rise (time interval A in Figure \ref{batse}b). Table~\ref{tab1} also shows that the period $P$ is longer near the impulsive phase maximum than during the impulsive phase rise. This property was seen also in other flares (see Table~1 in Paper II). We interpret the variations of $P$ as being due to variations of the length of oscillating magnetic traps (see Sections \ref{img} and \ref{disc}).

Low-amplitude fluctuations are seen after the impulsive phase (time interval B in Figure \ref{batse}b). Table~1 shows that these fluctuations also contain a weak quasi-periodic component.

The light curves in Figures \ref{lc-hxt} and \ref{batse}a can be divided into two components: the HXR pulses and a ``quasi-smooth'' component (emission below the pulses in the figures). The full amplitude, Amp $S$, of the $S$-function for a HXR pulse is a measure of the ratio of the pulse intensity to the quasi-smooth component; the mean values of Amp $S$ are given in Table~1. Our interpretation of the quasi-smooth component has been given in Paper II, where we explained it as being the result of superposition of the emission generated by many magnetic traps whose oscillations are shifted in phase.

\subsection{Investigation of the X-ray Images} \label{img}

\begin{figure}
\centerline{\includegraphics[width=0.8\textwidth]{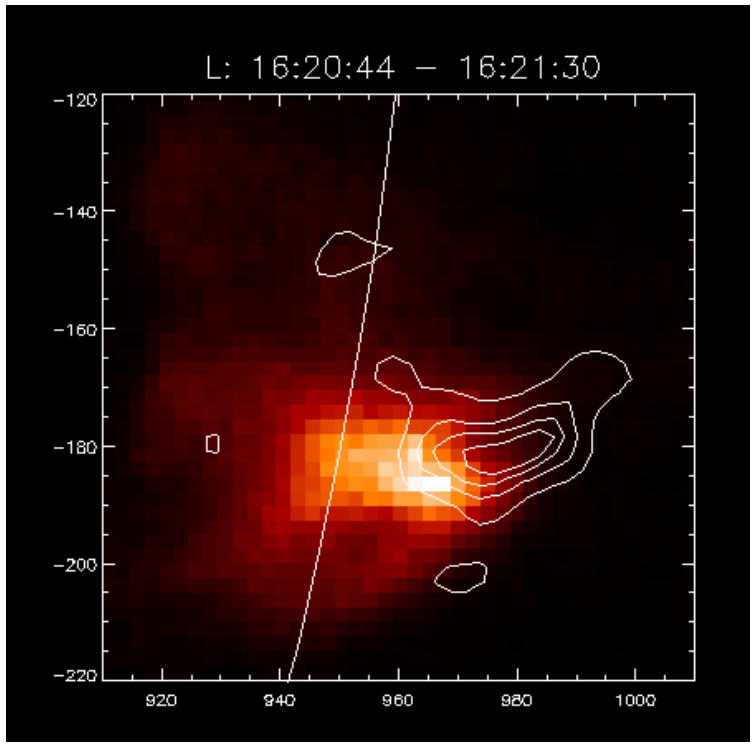}}
\hfill \caption{The {\sl Yohkoh} soft X-ray image recorded with the Be119 filter (the colour image) and the 14-23 keV HXR image (isocontours) at the beginning of the sequence of HXR pulses. The isocontours are 0.20, 0.39, 0.59, and 078$I_{{\rm max}}$, where $I_{{\rm max}}$ is the intensity of the brightest pixel.} \label{hxt1}
\end{figure}

\begin{figure}
\centerline{\includegraphics[width=1\textwidth]{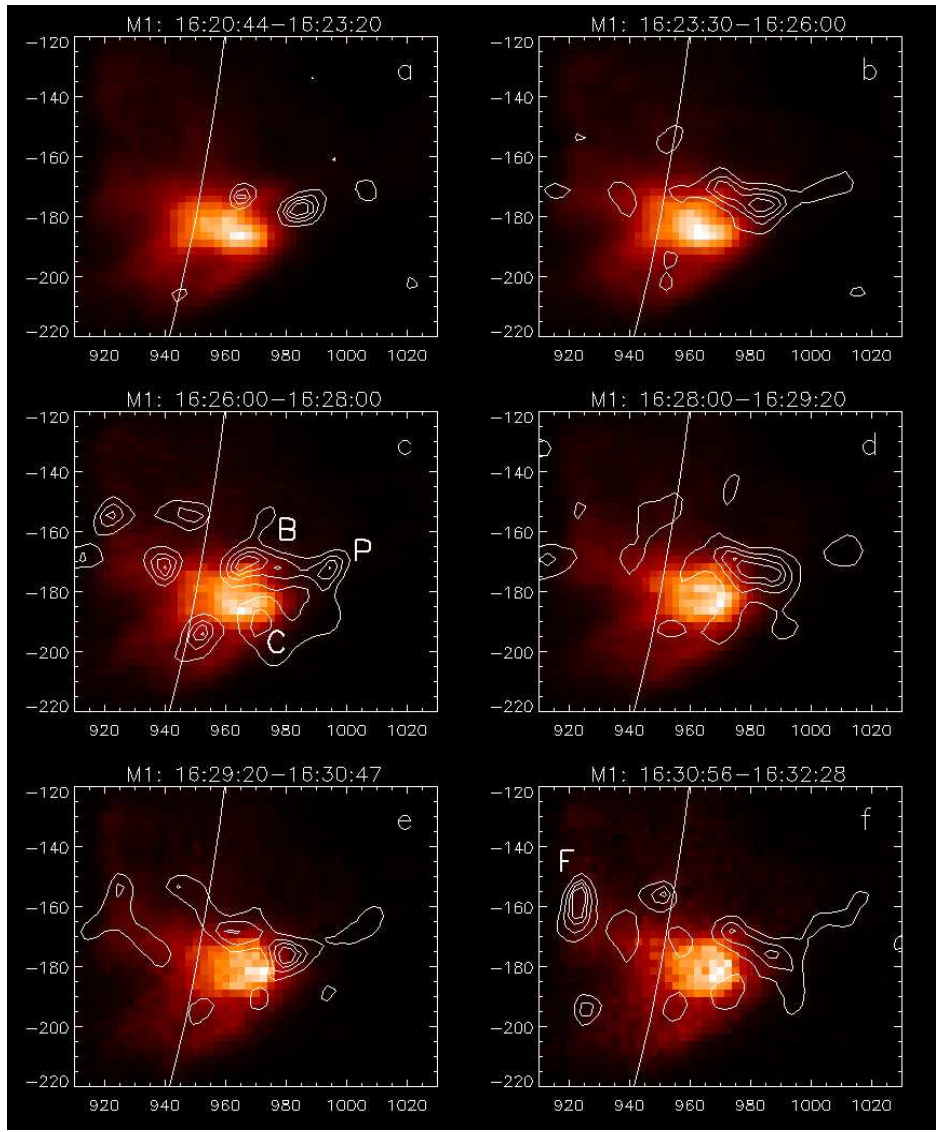}}
\hfill \caption{The sequence of SXR and 23-33 keV HXR images for the time interval 16:20--16:32 UT.} \label{hxt2}
\end{figure}

\begin{figure}
\centerline{\includegraphics[width=0.8\textwidth]{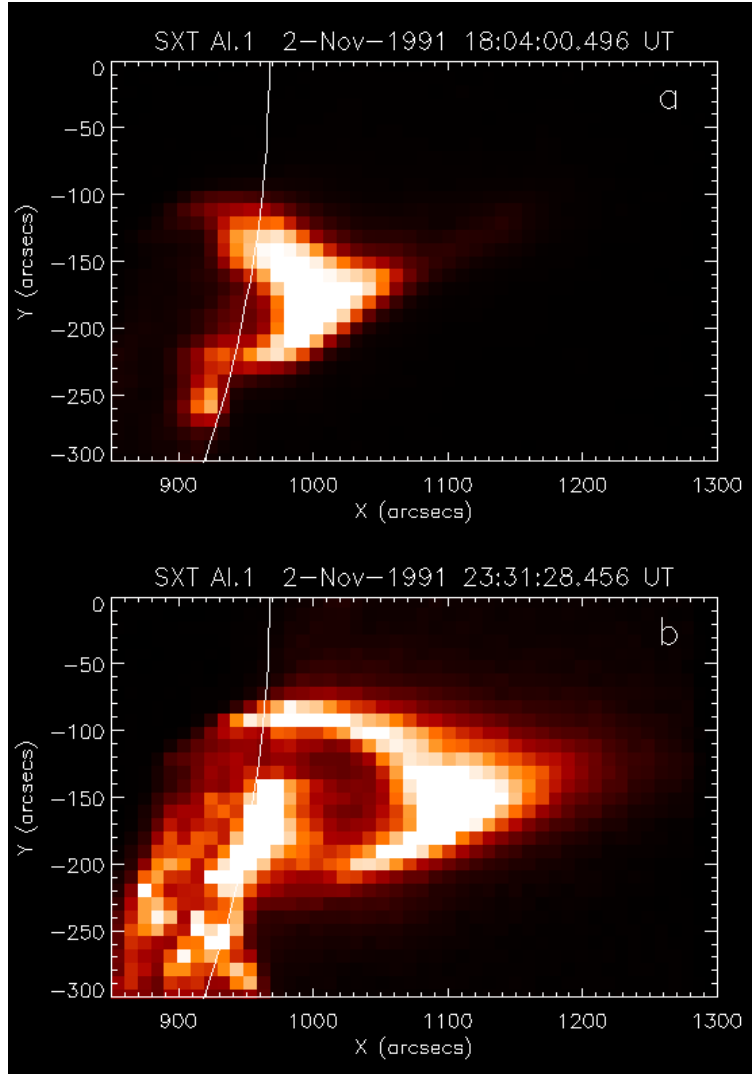}}
\hfill \caption{The late-phase SXR images of the flaring loop taken with the Al.1 filter The cusp-like structure is clearly seen at the top of the loop.} \label{sxt}
\end{figure}

\begin{figure}
\centerline{\includegraphics[width=1\textwidth]{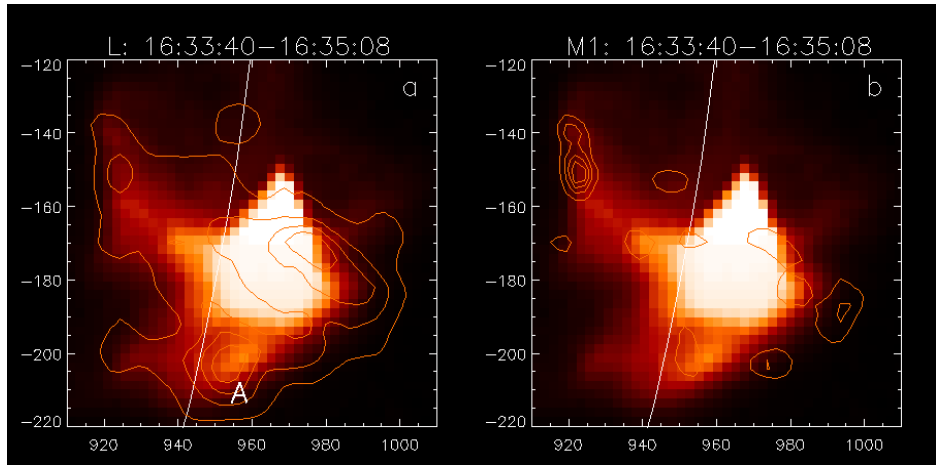}}
\hfill \caption{{\sl Yohkoh} soft-X-ray image for 16:34:08 UT (intensity scale in colour). HXR images are taken at the HXR maximum: a) {\sl Yohkoh} 14-23 keV image for 16:33:40-16:35:08 UT (isocontours). b) {\sl Yohkoh} 23-33 keV image for 16:33:40-16:35:08 UT (isocontours).} \label{hxt3}
\end{figure}

\begin{figure}
\centerline{\includegraphics[width=1\textwidth]{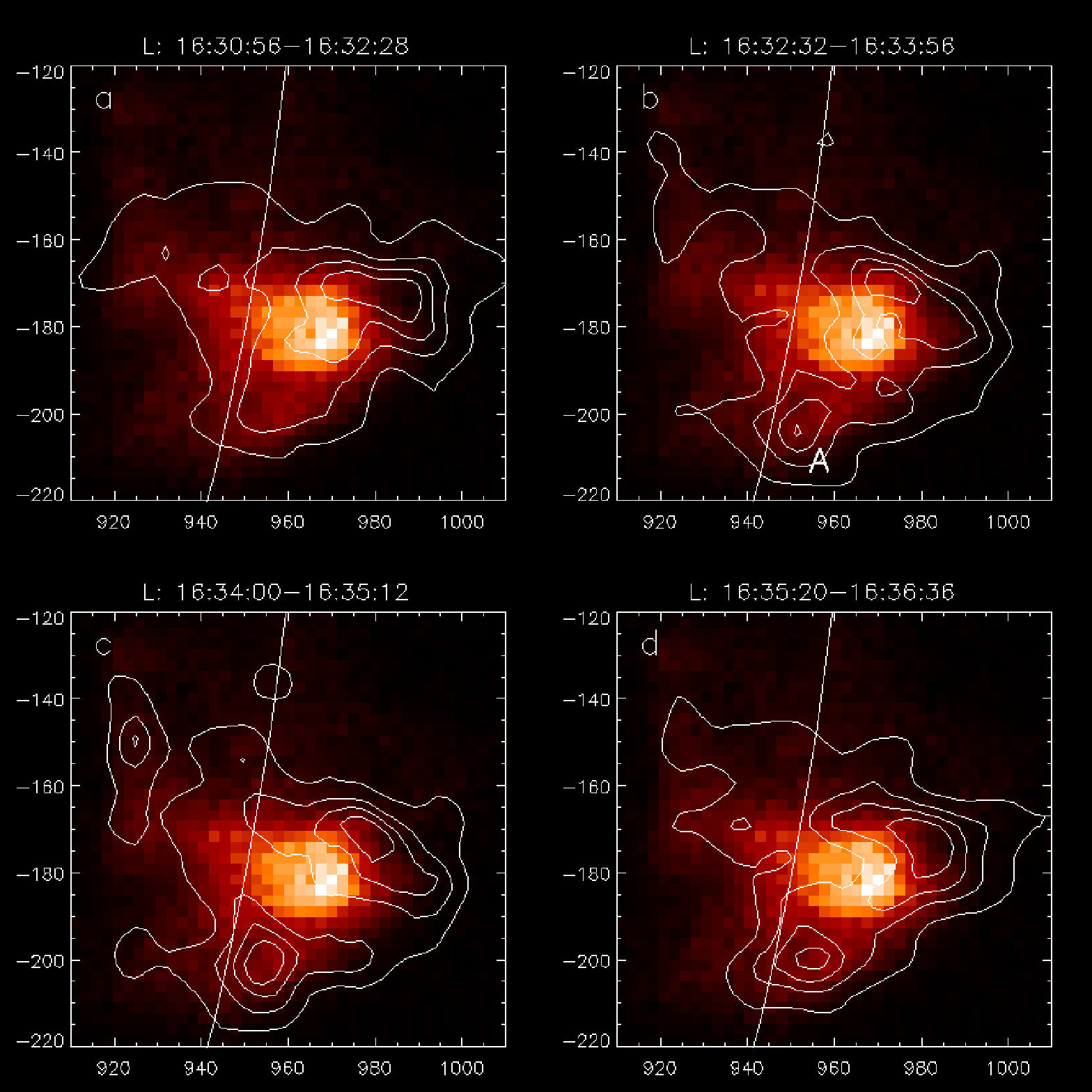}}
\hfill \caption{The sequence of 14-23 keV images for the time interval 16:30--16:36 UT. The same SXR image of 16:30:42 UT is displayed in all the images to avoid saturated images.} \label{hxt4}
\end{figure}

X-ray images of the flare of 2 November 1991 are shown in Figures \ref{hxt1}-\ref{sxt} and \ref{hxt3}-\ref{hxt4}. The color SXR images show a stable flaring loop. The bright central SXR source is the arcade channel which in this flare was perpendicular to the plane of the images.

\begin{figure}
\centerline{\includegraphics[width=0.8\textwidth]{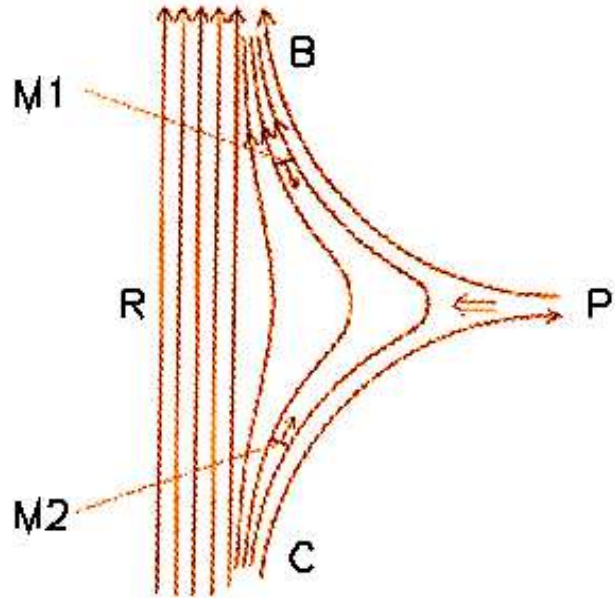}}
\hfill \caption{Magnetic structure of the BPC cusp. [Note: In the flare investigated the magnetic field R was perpendicular to the plane of images (it was the arcade channel)]. See text for details.} \label{triangle}
\end{figure}

The HXR images are displayed as isocontours. The sensitivity of the {\sl Yohkoh} Hard X-ray Telescope was moderate and the counting rates were low in this flare (see Figure~\ref{lc-hxt}); therefore it was necessary to apply rather long integration time in the HXR image reconstruction.

Figure~\ref{hxt1} shows a 14-23 keV (channel L) image at the beginning of the sequence of HXR pulses. It shows that most of the 15 keV electrons were confined within the loop-top source (no significant escape toward the footpoints).

Further development of the HXR source can be better seen in 23-33 keV (channel M1) images. Figure~\ref{hxt2} shows a sequence of the images. In Figure~\ref{hxt2}c is seen a triangular (``cusp'') structure BPC which is analogous to that in Figure~\ref{sep30}. The source P is at the top of the cusp and sources B and C are located at the places where the magnetic lines coming from the cusp meet the arcade channel. The B and C sources indicate that there is a connection between these magnetic lines and the magnetic field of the arcade channel, which allows the precipitating electrons to penetrate into the dense plasma of the channel, emit HXRs, and heat the channel. Hence, the B and C sources are the places where the magnetic traps are connected with the arcade channel (they are also the nodes of the trap oscillations). In Figures~\ref{hxt2}b and \ref{hxt2}d the sources B and P are connected (not resolved).

The cusp is directed almost radially, with a slight inclination to the north. This radial direction of the cusp has been confirmed by late-phase SXR observations (see Figure~\ref{sxt}).

A specific feature of the present flare is that in most images in Figure~\ref{hxt2} the HXR source B is much stronger than the source C. This indicates that the magnetic field within the cusp was asymmetric and the asymmetry could change in time (see Section~\ref{disc} for discussion).

Figure~\ref{triangle} shows the structure of the magnetic field within the cusp (see Papers I and II). The structure is similar to that proposed by \citeauthor{asc04a} (\citeyear{asc04a}, \citeyear{asc04b}) and \inlinecite{asc96}. The magnetic reconnection is supposed to occur above the cusp P and the reconnection outflow excites the oscillations of the magnetic traps within the cusp.

Figure~\ref{hxt2} also shows that the precipitation of the 25 keV electrons toward the loop footpoints evolved in time. In Figures \ref{hxt2}a, \ref{hxt2}b, \ref{hxt2}d, and \ref{hxt2}e the precipitation is weak (weak footpoint emission). In Figure \ref{hxt2}c the sources near the footpoints are strong which means strong precipitation. About 16:29 UT the footpoint source F began to grow in intensity and after 16:31 UT it was dominant in the M1 images (see Figure~\ref{hxt2}f) which means that the precipitation was very efficient. This indicates that strong chromospheric evaporation had been initiated then. The changes of the precipitation of accelerated electrons from the BPC source will be discussed in Section~\ref{disc}. In Figure~\ref{hxt2}f we also see a strong asymmetry in the 23-33 keV footpoint emission which is analogous to the asymmetry between the B and C sources. This is also the result of asymmetry in the precipitation of 25 keV electrons from the cusp.

After 16:32 UT a strong increase in the SXR emission from the loop-top source occurred which is seen as the saturation of the SXR images (Figure~\ref{hxt3}), and about 16:34--16:35 UT a strong peak was seen in the HXR light curves (Figures \ref{lc-hxt} and \ref{batse}a). These two features are clearly due to the fact that dense plasma reached the loop-top and therefore a large number of electrons had been accelerated (see Section~\ref{disc}). These observations provide us direct evidence that the strong HXR pulses are the result of the inflow of dense plasma coming from the chromospheric evaporation, into the loop-top source.

Figure~\ref{hxt4} shows 14-23 keV (channel L) images taken at 16:31-16:37 UT. The sources B and P are connected (not resolved) and they remain strong in these images. The footpoint source F is strong only in Figure~\ref{hxt4}c. An additional source A developed in the southern ``leg'' of the flaring loop. This indicates that after 16:32 UT many 15-keV electrons were precipitated into the loop leg, where they met dense plasma coming from the chromospheric evaporation.

In Section~\ref{disc} we will show that the observed properties of the loop-top source can be explained in terms of our model of oscillating magnetic traps.

\section{Discussion} \label{disc}

In Papers I and II we have worked out the model of oscillating magnetic traps to explain the quasi-periodic oscillations of the HXR emission. This model is based on:
\begin{enumerate}[i)]
\item the magnetic structure of the cusp according to \inlinecite{asc96} and \inlinecite{asc04a};
\item the mechanism of electron acceleration in magnetic traps during their compression (\opencite{s+k97}; \opencite{k+k04});
\item our investigation of quasi-periodic variation in the HXR emission of solar flares (Papers I and II).
\end{enumerate}
According to our model the cusp (triangular) volume is filled with magnetic traps (see Figure~\ref{triangle}). Magnetic reconnection at the top of the cusp (near P) generates the reconnection outflow which excites the oscillations of the traps. Electrons are accelerated in the oscillating traps and they emit hard X-rays.

The non-decaying sequences of the HXR pulses, like 16:20-16:32 UT in Figures \ref{lc-hxt} and \ref{batse}, indicate that they are maintained by a feedback mechanism. Such a mechanism has been proposed in Paper II (the feedback between the amplitude of the oscillations of magnetic traps and the pressure of accelerated electrons).

In Figure~\ref{hxt2} we also see that the size of the BP source and the distribution of the HXR emission within the source changed in time. This is well explained by the model of oscillating magnetic traps, since: i) the BC magnetic mirrors are moving during the pulses (see Papers I and II), ii) the distribution of the plasma and accelerated electrons within the traps changes during the pulses.

Let us also note that the size and the distribution of the emission in the BC source are more stable in the L-band images (see Figures \ref{hxt3} and \ref{hxt4}). This is due to the fact that 15 keV electrons are generated in the traps whose maximum compression is weaker (larger $\chi_{{\rm min}}$ values, see below).

The sequences of the HXR images (Figures \ref{hxt2} and \ref{hxt3}-\ref{hxt4}) allowed us also to investigate time variation of the precipitation of the accelerated electrons from the magnetic traps. The precipitation of electrons from a magnetic trap is controlled by the trap ratio, $\chi = B_{{\rm max}}/B_{{\rm min}}$, where $B_{{\rm max}}$ is the magnetic field strength at the mirrors B and C, and $B_{{\rm min}}$ is the strength at the middle of the trap. In the oscillating magnetic traps the parameter $\chi$ decreases during the compression of the traps and reaches its minimum value, $\chi_{{\rm min}}$, during the maximum of compression, when the precipitation reaches its maximum.

The efficiency of particle precipitation steeply depends on the value of $\chi_{{\rm min}}$. To illustrate this dependence we have made some estimates under simplifying assumption that the distribution of accelerated electrons is isotropic. We describe the efficiency of precipitation by the ratio
\begin{equation}
r = m/n,
\end{equation}
where $m$ is the number (s$^{-1}$) of escaping electrons, and $n$ is the number (s$^{-1}$) of electrons which emit photons and are thermalized before they escape.

For the isotropic distribution
\begin{equation}
r = (1 - {\cos}{\alpha}_{\rm c})/{\cos}{\alpha}_{{\rm c}},
\end{equation}
where ${\alpha}_c$ is the critical value of the pitch angle measured at the middle of the trap, such that electrons with $\alpha < {\alpha}_{{\rm c}}$ escape. The value of ${\alpha}_{{\rm c}}$ for the maximum of compression is related to $\chi_{{\rm min}}$:
\begin{equation}
\chi_{min} = 1/{\sin}^2{\alpha}_c.
\end{equation}
Combining Equations (3) and (4) we obtain
\begin{equation}
r = \frac{1}{\sqrt{1 - 1/{\chi}_{{\rm min}}}} - 1.
\end{equation}
The function $r({\chi}_{{\rm min}})$ is given in Table~\ref{tab2}. The table shows that when we observe strong precipitation ($r > 1$), this means a very strong compression of the traps ($\chi_{{\rm min}} < 1.3$).

\begin{table}
\caption{The function $r({\chi}_{{\rm min}})$}
\label{tab2}
\begin{tabular}{ccccccc}
\hline
${\chi}_{{\rm min}}$ & 1 & 1.33 & 2 & 3 & 4 & 5 \\
\hline
$r$ & $\infty$ & 1.00 & 0.41 & 0.22 & 0.15 & 0.12 \\
 \hline
\end{tabular}
\begin{list}{}{}
\item $\chi_{{\rm min}}$ is the trap ratio at the maximum of compression.
\item $r = m/n$, where $m$ is the number (s$^{-1}$) of escaping electrons, \\ $n$ is the number (s$^{-1}$) of electrons which emit photons and \\ are thermalized before they escape.
\end{list}
\end{table}

Figure~\ref{hxt2} in connection with Table~\ref{tab2} indicates that during the sequence of pulses, 16:20-16:29 UT, the values of $\chi_{{\rm min}}$ for the 25 keV electrons were $\chi_{{\rm min}} > 3$ (weak precipitation of the electrons) and after 16:31 UT they were $\chi_{{\rm min}} < 1.3$ (strong precipitation). This shows that the dramatic change in the precipitation can be caused by a moderate change in the value of $\chi_{{\rm min}}$.

According to our model of oscillating magnetic traps the decrease in the value of $\chi_{{\rm min}}$ in the sequence of pulses ({\sl i.e.} increase in the maximum compression) is due to the increase in the amplitude of the magnetic trap oscillation. The amplitude gradually increases due to the feedback with the pulses of the pressure of accelerated (non-thermal) electrons.

When the precipitation of 25 keV electrons began to be strong (after 16:31 UT), the strong footpoint source F was seen in the M1 images (Figure~\ref{hxt2}). This shows that strong chromospheric evaporation began then. The dense plasma reached the loop-top about 16:32 UT which caused i) the strong SXR emission seen as the saturation in the SXR images (Figure~\ref{hxt3}) and ii) the strong HXR peak at about 16:34 UT (Figure~\ref{batse}a) which is due to the increase in the density in the oscillating magnetic traps. Figure~\ref{hxt3}b shows that strong precipitation of 25 keV electrons toward the footpoint F continued during the HXR maximum.

Further increase in the density inside the traps ($N\approx10^{11}$~cm$^{-3}$) caused damping of electron acceleration when the collisional energy loss time, ${\tau}_{\rm c}$, became similar to the acceleration time, ${\tau}_{{\rm acc}}$. This is seen as the quick decrease in HXR emission after 16:36 UT (Figure~\ref{batse}a).

Another important feature of the flare investigated is the strong asymmetry in electron precipitation (source B was much stronger than source C, and the northern footpoint source F dominated over southern footpoint). This asymmetry is the result of inclination of the cusp structure toward north (see Figure~\ref{hxt2}c). Therefore the oscillations of the traps were asymmetric: the compression near the B source was stronger (${\chi}_{{\rm min}}$ lower) than near the C source and hence the precipitation from B was much stronger than from C.

 Next property seen in the X-ray images (Figures \ref{hxt2} and \ref{hxt3}-\ref{hxt4}) is that after 16:31 UT the 25 keV electrons (M1 images) could easily escape from the BP source, but the precipitation of 15 keV electrons (L images) remained low. This indicates that in the ensemble of the traps which generated the accelerated electrons, traps had different values of $\chi_{{\rm min}}$.

The comparison between the L and M1 images (Figures \ref{hxt2} and \ref{hxt3}-\ref{hxt4}) also shows that the BP source was systematically larger in the L images than in the M1 images. This shows that the traps which generated the 15 keV electrons occupied a larger volume in the BP source than the traps which generated the 25 keV electrons.

\begin{figure}
\centerline{\includegraphics[width=1\textwidth]{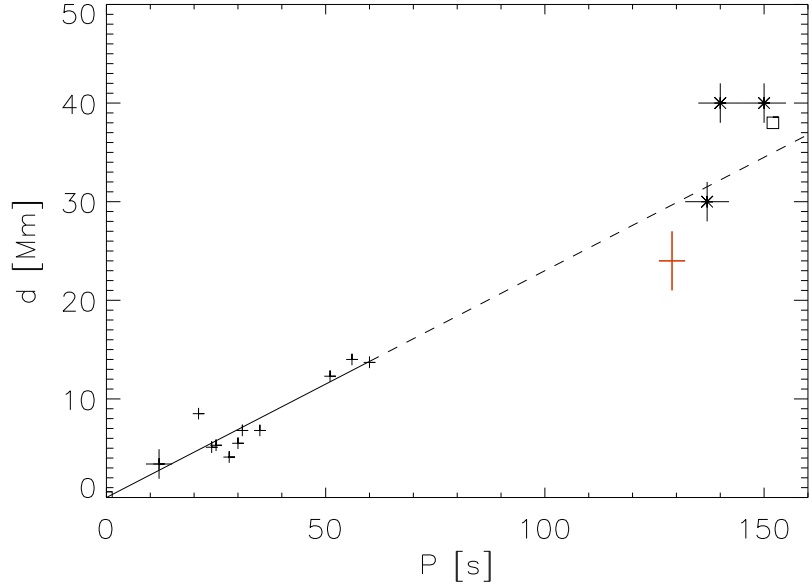}}
\hfill \caption{The relationship between the size, $d$, of the HXR loop-top source and the period, $P$, obtained in Paper I. The regression line has been calculated for smaller flares ($P < 60$~s). The thick red cross shows the position of the flare investigated in the present paper.}
\label{plot}
\end{figure}

In Paper I we investigated the relationship between the period and the size of the HXR loop-top sources. Figure~\ref{plot} shows the obtained correlation diagram. In that paper we described the size of the BPC cusp in large flares with a parameter $2a$, where $a$ is the height of the triangle BPC. Therefore for the present flare we have measured the value $a \approx 12$ Mm (Figure~\ref{hxt2}c) and plotted the point ($P$, $2a$) as the thick red cross in Figure~\ref{plot}.

It may be rather surprising that in Figure~\ref{plot} the points corresponding to the large ($P > 120$~s) and small ($P < 60$~s) flares are located near the same regression line. This can be easily explained in terms of the model of oscillating magnetic traps.
\begin{enumerate}[i)]
\item In Paper I we have obtained the following estimate:
\begin{equation}
d/P \approx (1/\pi)v_{\rm A},
\end{equation}
where $v_{\rm A}$ is the Alfv$\rm{\acute{e}}$n speed in the magnetic traps.
\item In Paper I we have also found that during the impulsive phase the plasma pressure, $p$, inside the traps reaches values which are of the order of the magnetic pressure
\begin{equation}
p \approx B^2/(8\pi)
\end{equation}
(see Section~3 in Paper I).
\end{enumerate}

Using the relationships:
\begin{equation}
p = 2NkT,
\end{equation}
\begin{equation}
v_{\rm A} = B/\sqrt{4\pi\rho},
\end{equation}
where $N$ is the electron number density, $k$ is the Boltzmann constant, and $\rho$ is the plasma density, we obtain:
\begin{equation}
v_{\rm A} \approx 1.8 \times 10^4 \sqrt{T} \quad\mbox{[cm\,s$^{-1}$]}\quad
\end{equation}
From Equations (6) and (10) we obtain:
\begin{equation}
d/P \approx 5.7 \times 10^3 \sqrt{T} \quad\mbox{[cm\,s$^{-1}$]}\quad
\end{equation}
Hence the points corresponding to large and small flares are located near the same regression line because:
\begin{enumerate}[i)]
\item the gas pressure, $p$, inside the magnetic traps is proportional to the magnetic pressure [Equation (7)],
\item the temperatures which are reached in these flares are similar [see Equation (11)].
\end{enumerate}

We also wanted to check if Equation (11) provides good approximation of the $d/P$ ratio. Therefore we have estimated the temperature, $T$, and the electron number density, $N$, in the BP source from the SXR images which have been recorded with the Be119 and Al.1 filters.  This diagnostics was only possible for the time interval 16:20-16:30 UT, since later images were seriously saturated (see Figure~\ref{hxt3}). Using the filter ratio method we have found that the temperature slowly increased from 11 to 12 MK and the electron number density was about $2 \times 10^{10}$~cm$^{-3}$ during this time interval.

\begin{figure}
\centerline{\includegraphics[width=1\textwidth]{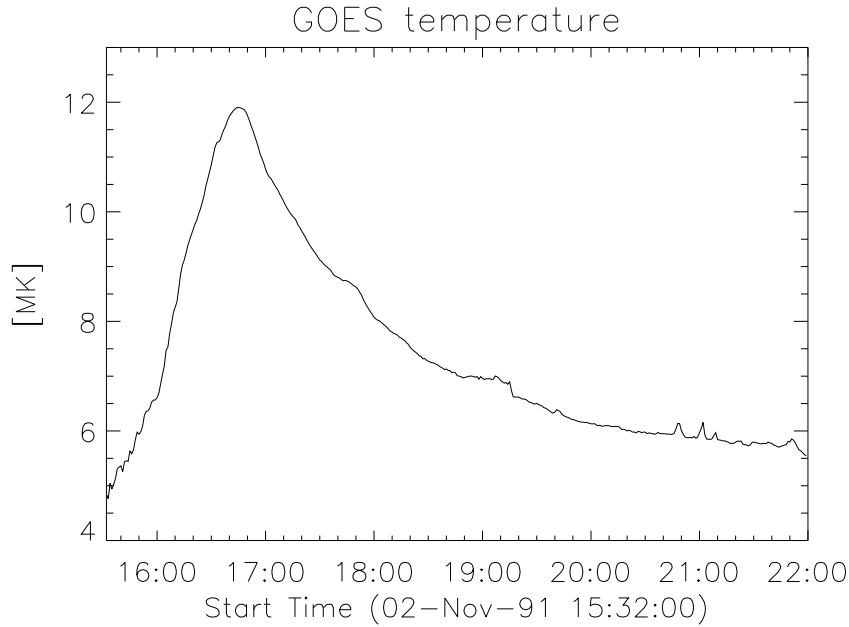}}
\hfill \caption{Time variation of the temperature, $T_{{\rm GOES}}$, derived from the GOES observations.}
\label{tgoes}
\end{figure}

It has been shown \cite{j+b11} that the temperature, $T_{{\rm GOES}}$, estimated from the standard GOES records, adequately represents the temperature of the hot plasma which generates the investigated SXR emission. The obtained values of $T_{{\rm GOES}}$ are shown in Figure~\ref{tgoes}. We see that during the impulsive phase the temperature reached 12 MK which confirms that the flare was not very hot. This is in agreement with the temperatures $T_{{\rm GOES}}$ obtained for other ``slow-LDE'' flares \cite{b+j05}.

Putting the value $T = 12$~MK into Equation (11) we obtain $d/P \approx 0.20$ Mm\,s$^{-1}$ which is in good agreement with the value obtained from the observations (Figure~\ref{plot}), $d/P = 0.18 \pm 0.04$ Mm\,s$^{-1}$. Hence Equation (11) adequately describes the investigated flare and it probably provides a good approximation for other flares which are not very hot.

Figure~\ref{sxt} shows late-phase SXR images of the investigated flare. These images are very important to confirm the cusp-like magnetic structure of the flare. A slow rise of the cusp structure is seen with the velocity about 4 km\,s$^{-1}$.

\section{Summary} \label{sum}

The flare investigated in the present paper is exceptional, because:
 \begin{enumerate}[i)]
 \item it was a large slowly-developing flare located at the solar limb,
 \item it showed a clear sequence of quasi-periodic HXR pulses,
 \item the available X-ray observations of the flare allowed us to investigate the development of its impulsive phase in some details.
 \end{enumerate}
The main aim of the present paper was to show that the observations of the quasi-periodic oscillations in the HXR emission contain important information about the electron acceleration.

The analysis of the HXR observations has confirmed that i) quasi-periodic oscillations of the magnetic traps occur in the cusp-like magnetic structure and ii) the oscillations accelerate the electrons which generate the quasi-periodic HXR pulses.

The analysis of the observations allowed us also to investigate changes in the efficiency of precipitation of accelerated electrons during the impulsive phase. Moreover, we have found that the efficiency of precipitation ({\sl i.e.} $\chi_{{\rm min}}$) is different in different traps which participate in the generation of the HXR pulses.

Other important findings are the following:
\begin{enumerate}[i)]
\item The HXR pulses increase their amplitudes due to the increase in the amplitude of magnetic trap oscillations and the increase in the density within the traps caused by the chromospheric evaporation upflow.
\item Accelerated electrons can precipitate into the arcade channel.
\item The increase in the amplitude of the HXR pulses terminates when further increase in the density inside the traps inhibits the acceleration of electrons.
\end{enumerate}

Generally, the model of oscillating magnetic traps is able to explain all the properties which have been derived from the analysis of the HXR emission in the present paper. In particular, the model explains: i) time variation of the electron precipitation seen in the sequences of the HXR images, ii) strong asymmetry in the precipitation of accelerated electrons, and iii) systematic differences in the precipitation of 15 and 25 keV electrons revealed by the comparison of the L and M1 images. All this means that the HXR observations, together with the model of oscillating magnetic traps, give us a consistent picture of the development of the flare impulsive phase.

\begin{acks}
The {\sl Yohkoh} satellite is a project of the Institute of Space and Astronautical Science of Japan. The {\sl Compton Gamma Ray Observatory} is a project of NASA. The authors are very thankful to the anonymous referee for her/his important remarks which helped to improve this paper. We acknowledge financial support from the Polish National Science Centre grant 2011/03/B/ST9/00104.
\end{acks}

\end{article}

\end{document}